\documentclass[3p,twocolumn,10pt]{elsarticle}
\biboptions{sort&compress}
\usepackage{amsmath}
\usepackage{graphicx}
\usepackage{bm}
\usepackage{slashed}
\usepackage{amssymb}
\usepackage{xcolor}
\usepackage{verbatim}
\usepackage{todonotes}
\usepackage{grffile}
\usepackage{soul}

\begin{document}

\begin{frontmatter}

\author[sut]{Christoph Herold}
\ead{herold@g.sut.ac.th}
\author[sut]{Apiwit Kittiratpattana}
\author[sut]{Chinorat Kobdaj\corref{cor1}}
\ead{kobdaj@g.sut.ac.th}
\author[sut]{Ayut Limphirat}
\author[sut]{Yupeng Yan}
\author[nan]{Marlene Nahrgang}
\author[fia]{Jan Steinheimer}
\author[fia,the,gsi,jul]{Marcus Bleicher}
\address[sut]{School of Physics, Suranaree University of Technology, 111 University Avenue, Nakhon Ratchasima 30000, Thailand}
\address[nan]{SUBATECH, UMR 6457, Universit́e de Nantes, IMT Atlantique, IN2P3/CNRS. 4 rue Alfred Kastler, 44307 Nantes cedex 3, France}
\address[fia]{Frankfurt Institute for Advanced Studies, Ruth-Moufang-Str. 1, 60438 Frankfurt am Main, Germany}
\address[the]{Institut f¨ur Theoretische Physik, Goethe Universit¨at Frankfurt,
Max-von-Laue-Strasse 1, D-60438 Frankfurt am Main, Germany}
\address[gsi]{GSI Helmholtzzentrum f¨ur Schwerionenforschung GmbH, Planckstr. 1, 64291 Darmstadt , Germany and}
\address[jul]{John von Neumann-Institut f¨ur Computing, Forschungzentrum J¨ulich, 52425 J¨ulich, Germany}

\cortext[cor1]{Corresponding author}

\title{Entropy production and reheating at the chiral phase transition}

\begin{abstract}
We study the production of entropy in the context of a nonequilibrium chiral phase transition. The dynamical symmetry breaking is modeled by a Langevin equation for the order parameter coupled to the Bjorken dynamics of a quark plasma. We investigate the impact of dissipation and noise on the entropy and explore the possibility of reheating for crossover and first-order phase transitions, depending on the expansion rate of the fluid. The relative increase in $S/N$ is estimated to range from $10\%$ for a crossover to $200\%$ for a first-order phase transition at low beam energies, which could be detected in the pion-to-proton ratio as a function of beam energy.
\end{abstract}


\end{frontmatter}

\section{Introduction}

Ultrarelativistic heavy-ion collsions experiments at RHIC and LHC have provided strong evidence for a quark-gluon plasma (QGP) phase at large temperatures and densities. One of the characteristics of this new phase of strongly-interacting matter is the restoration of chiral symmetry which is spontaneously broken in the ground state of quantum chromodynamics (QCD). Lattice QCD studies have revealed a crossover chiral transition for small baryochemical potentials $\mu_{\rm B}$ \cite{Aoki:2006we,Borsanyi:2010bp,Borsanyi:2013bia,Bazavov:2014pvz}, while evidence for a critical point (CP) and first-order phase transition has been obtained from functional methods allowing the exploration of regions with large values of $\mu_B$ \cite{Fischer:2012vc,Gao:2016qkh}. Recently, considerable progress has been made in constructing a QCD equation of state with a CP based on lattice QCD data \cite{Parotto:2018pwx}.

The experimental search for the QCD phase transition is nowadays one of the central goals for current and future collider facilities. It is mainly driven by measurements of net-proton or net-charge multiplicity fluctuations \cite{Aggarwal:2010wy,Luo:2012kja,Adamczyk:2013dal,Luo:2015ewa} which are expected to show characteristic nonmonotonic behavior near the phase transition and especially near the CP where the correlation length diverges \cite{Asakawa:2000wh,Stephanov:1998dy,Stephanov:1999zu,Cheng:2008zh,Asakawa:2009aj,Gupta:2011wh,Friman:2011pf,Stephanov:2008qz,Athanasiou:2010kw}. A proper understanding of the experimental results, however, requires careful analysis of the dynamical processes that will have an influence especially near a CP where critical slowing down severely limits the growth of the correlation length and therefore of critical fluctuations \cite{Berdnikov:1999ph}, and memory effects can lead to a broadening of the critical region \cite{Herold:2017day}. Nonequilibrium models have tried to address these issues demonstrating critical slowing down near a CP \cite{Nahrgang:2011mv,Herold:2013bi,Mukherjee:2015swa,Stephanov:2017wlw,Nahrgang:2018afz} and spinodal decomposition resulting in domain formation at a first-order phase transition \cite{Randrup:2009gp,Randrup:2010ax,Steinheimer:2012gc,Herold:2013qda,Herold:2014zoa}. Near a CP, the nonequilibrium dynamics will influence the order parameter and, related to that, experimental observables such as net-baryon or net-proton multiplicity fluctuations \cite{Athanasiou:2010kw,Stephanov:2011pb,Herold:2016uvv}. 

While ideal hydrodynamics preserves the total entropy, the addition of dissipation and fluctuation in nonequilibrium models will lead to an increasing entropy, an effect that is going to be addressed in this paper both qualitatively and quantitatively, together with its potential as an experimental signal for a first-order QCD phase transition. One can expect that a delay in the relaxation of the critical mode produces additional entropy \cite{Csernai:1992as}. In this work, we explore the impact of dynamical effects on the production of entropy within a model describing the realistic dynamics of the chiral transition, the recently studied Nonequilibrium Chiral Fluid Dynamics (N$\chi$FD) \cite{Mishustin:1998yc,Nahrgang:2011mg}. We investigate the evolution of the critical $k=0$ mode of the chiral order parameter field coupled to the longitudinal Bjorken-type expansion of a quark-antiquark fluid. It should be noted that this approach does not cover various other sources of entropy production such as shear viscosity \cite{Dumitru:2007qr} or compression during the early stage of a heavy-ion collision \cite{Reiter:1998uq}.

Results presented in this article are of considerable interest for experiments at the future facilities of FAIR \cite{Friman:2011zz}, NICA \cite{nica:whitepaper}, and also for RHIC's BES II program. 

The present paper is structured as follows: We present the dynamical equations of our model in Sec.\ \ref{sec:model} and apply these in Sec.\ \ref{sec:dissfluct} to study the production of entropy from dissipation and fluctuations. Sec.\ \ref{sec:expansion} then presents results on the impact of initial conditions on the entropy increase and reheating for crossover and first-order phase transition evolutions. This will be further investigated in Sec.\ \ref{sec:entropy} where these effects are compared for different transition scenarios. In the end, we will conclude with a summary and outlook in Sec.\ \ref{sec:conclusions}.

\section{Chiral Bjorken dynamics}
\label{sec:model}

Our ansatz for the chiral phase transition is the linear sigma or quark-meson model which has the essential features of a crossover for small baryochemical potentials and a CP and first-order phase transition at large values of $\mu_{\rm B}$. The Lagrangian density of this model is
\begin{align}
\label{eq:Lagrangian}
 {\cal L}&=\overline{q}\left(i \gamma^\mu \partial_\mu-g \sigma\right)q + \frac{1}{2}\left(\partial_\mu\sigma\right)^2- U(\sigma)~, \\
 U(\sigma)&=\frac{\lambda^2}{4}\left(\sigma^2-f_{\pi}^2\right)^2-f_{\pi}m_{\pi}^2\sigma +U_0~.    
\end{align}
Here, we have readily set the pion fields equal to their vacuum expectation value of zero as we focus on the evolution of the field $\sigma$ as the chiral order parameter. The field $q=(u,d)$ includes the light quark fields only. The parameters of this model are chosen in the standard fashion with $f_\pi=93$~MeV, $m_\pi=138$~MeV and $U_0$ such that the potential $U(\sigma)$ vanishes in the ground state. The quark-meson coupling $g$ is fixed by the condition that $g\sigma$ equals the nucleon mass of around $940$~MeV in vacuum. 

The equation of motion for the zero mode $\sigma(\tau)=\frac{1}{V}\int\mathrm d^3 x\sigma(\tau,x)$ reads
\begin{equation}
 \label{eq:eom_sigma}
 \ddot\sigma+\left(\frac{D}{\tau}+\eta\right)\dot\sigma+\frac{\delta\Omega}{\delta\sigma}=\xi~,
\end{equation}
with the dot referring to the derivative with respect to proper time $\tau$. The potential $\Omega=U+\Omega_{q\bar q}$ contains the mean-field quark-antiquark contribution
\begin{align}
 \Omega_{q\bar q}=-2N_f N_c T\int \frac{\mathrm d^3 p}{(2\pi)^3} & \left[\log\left(1+\mathrm e^{-\frac{E-\mu}{T}}\right)\right. \\ & \left. +\log\left(1+\mathrm e^{-\frac{E+\mu}{T}}\right)\right]~, 
\end{align}
with $N_f=2$, $N_c=3$ being the number of light quark flavors and colors, $T$ the temperature and $\mu=\mu_B/3$ the quark chemical potential. The dynamically generated energy of a constituent quark with momentum $p$ is $E=\sqrt{p^2+g^2\sigma^2}$. 
In the Hubble term $\sim D/\tau$, we set $D=1$, considering the case of a longitudinal expansion along the direction of the beam axis. The damping coefficient $\eta$ describes various dissipative processes of the sigma field: First, mesonic interactions, i.e.\ scattering of a condensed sigma meson with a thermal sigma, $\sigma\sigma\leftrightarrow\sigma\sigma$, and two-pion decay, $\sigma\leftrightarrow\pi\pi$ \cite{Csernai:1999ca}. Second, meson-quark interactions, $\sigma\leftrightarrow q\bar q$ \cite{Nahrgang:2011mg}. We include all of these within a phenomenological constant damping coefficient of $\eta=2.2/$fm \cite{Biro:1997va}. The stochastic noise field $\xi$ is assumed to be white and Gaussian, characterized by mean and variance, 
\begin{align}
 \langle\xi(t)\rangle&=0~,\\
 \langle\xi(t)\xi(t')\rangle&=\frac{2T\eta}{V}\delta(t-t')~.
\end{align}

We assume the quark degrees of freedom to constitute an ideal fluid with energy-momentum tensor $T^{\mu\nu}_q=(e+p)u^{\mu}u^{\nu}-pg^{\mu\nu}$. Energy-momentum conservation now dictates the vanishing of the divergence of the total energy-momentum tensor, 
\begin{equation}
 \partial_\mu T^{\mu\nu}=\partial_\mu \left(T^{\mu\nu}_q+T^{\mu\nu}_\sigma\right)=0~.
\end{equation}
A self-consistent derivation within the two-particle irreducible action formalism as in \cite{Nahrgang:2011mg} yields
\begin{equation}
\label{eq:eom_fluid}
 \partial_\mu T^{\mu\nu}_q=\left[\frac{\delta\Omega_{q \bar q}}{\delta\sigma}+\left(\frac{D}{\tau}+\eta\right)\dot\sigma\right]\partial^\nu \sigma.
\end{equation}
Contracting Eq.~\eqref{eq:eom_fluid} with the four-velocity $u^\nu$ gives the equation for the evolution of the energy density, 
\begin{equation}
\label{eq:eom_eden}
 \dot e=-\frac{e+p}{\tau}+\left[\frac{\delta\Omega_{q \bar q}}{\delta\sigma}+\left(\frac{D}{\tau}+\eta\right)\dot\sigma\right]\dot\sigma~.
\end{equation}
The net-baryon density follows the equation
\begin{equation}
 \label{eq:eom_nden}
 \dot n = -\frac{n}{\tau}~.
\end{equation}
Finally, the coupled set of equations \eqref{eq:eom_sigma} and \eqref{eq:eom_eden} is closed by the equation of state, $p=-\Omega_{q\bar q}$. The entropy density at each time $\tau$ is then given by 
\begin{equation}
 s=\frac{e+p-\mu n}{T}~,
\end{equation}
and is supposed to yield a conserved total entropy for ideal hydrodynamic evolution. In this case, the equations would need to be modified such that $e=T\frac{\partial p}{\partial T}-p+\mu n+U$ and $p=-\Omega_{q\bar q}-U$ together with the condition that the field $\sigma$ is equal to its equilibrium value at all times. This naturally leads to $s\tau=s_0\tau_0$. The evolution then simply follows the isentropes of the quark-meson model as they have been calculated in \cite{Scavenius:2000qd}, showing a characteristic bending of trajectories at the phase boundary. 

The Langevin equation \eqref{eq:eom_sigma} describes the relaxation of the critical mode. Near equilibrium, neglecting stochastic fluctuations, we may write this as
\begin{equation}
 \ddot\sigma+\eta\dot\sigma+m^2_\sigma(T)\sigma\approx 0~.
\end{equation}
Here, $m_\sigma(T)$ denotes the temperature-dependent screening mass of the sigma meson, defined as the second derivative of $\Omega$ with respect to $\sigma$ at its global minimum. The solution of this equation is given by $\sigma(\tau)\sim \mathrm e^{\alpha \tau}$ with 
\begin{equation}
 \alpha=-\frac{\eta}{2}\pm\sqrt{\frac{\eta^2}{4}-m^2_\sigma(T)}~. 
\end{equation}
For the underdamped or critically damped case with $m_\sigma(T)\geq \eta/2$, this gives a relaxation time of $2/\eta\approx0.9$~fm which is prolonged if $m_\sigma(T)< \eta/2$, which is the case around the transition and especially near the CP where $m_\sigma(T)\rightarrow 0$ and critical slowing down sets in.

\section{Entropy production from dissipation and fluctuations}
\label{sec:dissfluct}

\begin{figure}[t]
\includegraphics[width=0.6\textwidth,angle=270]{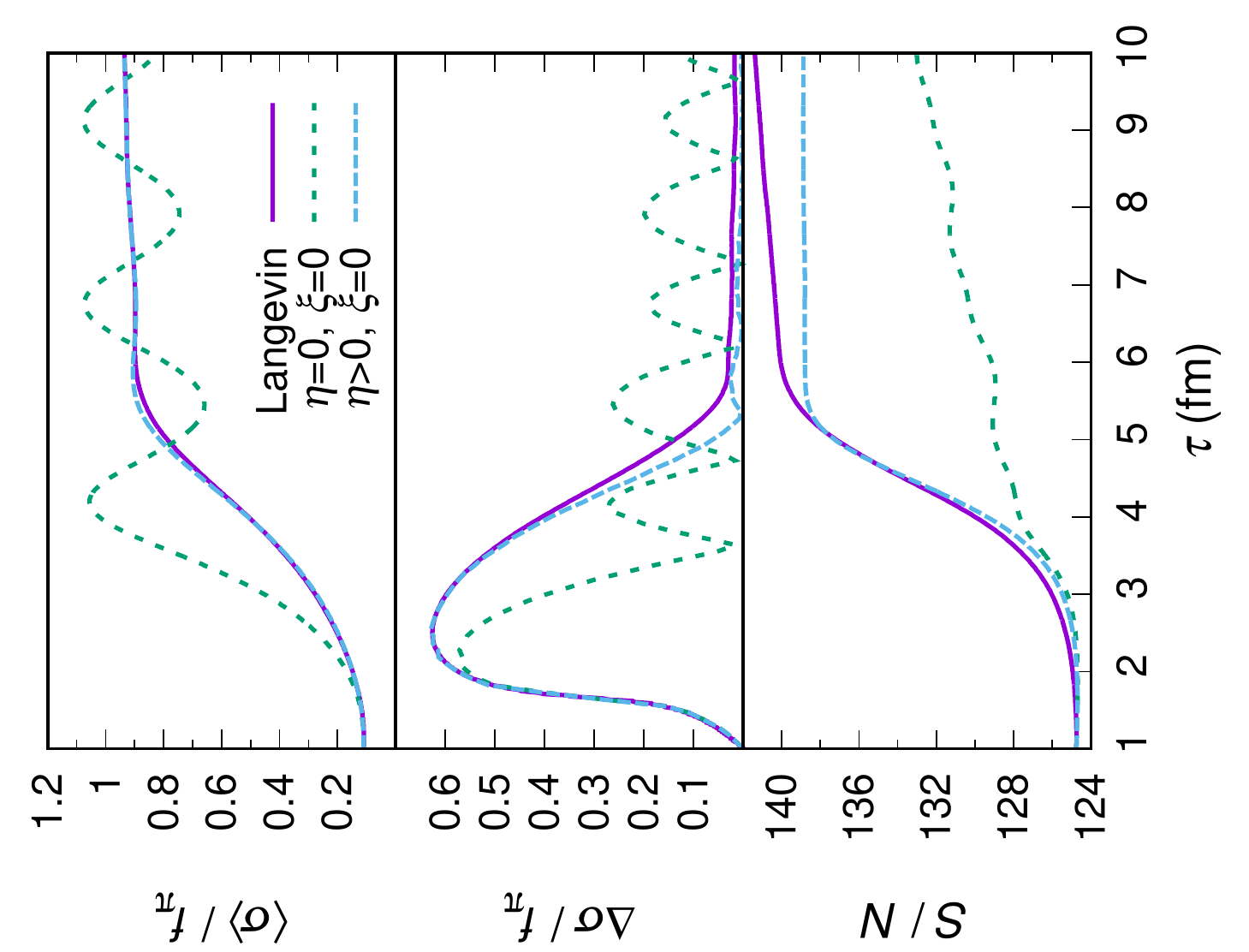}
\caption{\label{fig:sigma_t} Top: Event-averaged sigma field. Middle: Fluctuation of the sigma field from equilibrium $\Delta\sigma=\sqrt{\langle \sigma-\langle\sigma\rangle^2\rangle}$. Bottom: Entropy per baryon as a function of proper time $\tau$, comparing evolutions with a full Langevin dynamics, without dissipation and noise and with dissipation but no noise.} 
\end{figure} 

Besides the full Langevin dynamics, it is instructive to study Eq.~\eqref{eq:eom_sigma} without noise ($\xi=0$) and without dissipation and noise ($\eta=0$, $\xi=0$), the latter case representing a propagation according to the classical Euler-Lagrange equation. We choose an initial condition $(T_0,\mu_0)=(171,19)$~MeV, resulting in an evolution passing through the crossover region of the quark-meson model over a proper time from $\tau=1$ to $10$~fm. The initial entropy-to-baryon number ratio is $S/N=124.5$ and is constant for ideal hydrodynamics, characterized by the equilibrium condition $\sigma=\sigma_{\rm eq}$, with
\begin{equation}
 \frac{\partial\Omega}{\partial\sigma}\bigg|_{\sigma=\sigma_{\rm eq}}=0~. 
\end{equation}
The dynamical treatment of the chiral order parameter, however, raises the expectation that relaxational effects and fluctuations produce entropy during the evolution through and beyond the transition. We show the evolution of $\langle\sigma\rangle$, $\Delta\sigma=\sqrt{\langle (\sigma-\langle\sigma\rangle)^2\rangle}$ and $S/N$ as a function of $\tau$ in Fig.~\ref{fig:sigma_t}. Here, $\langle\sigma\rangle$ and $\Delta\sigma$ are scaled by the vacuum expectation value $f_\pi$ and $\langle\cdot\rangle$ denotes the event-average over different noise configurations which is applied in the case of the full Langevin dynamics. In the upper plot, we can follow the relaxation of the order parameter to equilibrium during the crossover transition. While the scenario without dissipation and noise leads to unphysical fluctuations around equilibrium, the damping term ensures the proper relaxational dynamics. The further addition of a noise term slightly increases the relaxation time, similar to what has been observed in \cite{Fraga:2007gg} for the Langevin dynamics of the SU(2) deconfining transition. In the behavior of $\Delta\sigma$, we see that the noise also prevents the fluctuations around the equilibrium value to vanish, ensuring that $\Delta\sigma$ remains finite also for later times, when fluctuations in the noise-free case with $\eta>0$ have already vanished. The initial rapid increase in $\Delta\sigma$ results form the rapid decrease in temperature which the order parameter is not able to follow immediately. The subsequent decay of this fluctuation from $\tau=2.5$ to $5.5$~fm evolves parallel to the rapid increase of $\sigma$ toward its low-temperature equilibrium value. In the same time, a clear increase in the entropy is seen for the two scenarios with $\eta>0$, as a result of the energy transfer $\sim\eta\dot\sigma$ during the rapid decay of the chirally restored phase. Besides the impact of friction, the plot on the bottom reveals another source of entropy production, namely fluctuations in the order parameter. For the full Langevin dynamics, these seem to produce a steady increase in entropy both before and after the transition process while the large fluctuations around the expectation value in the case without damping and noise leads to a stronger increase due to fluctuations during and after the transition.

\section{Impact of the expansion rate}
\label{sec:expansion}

\begin{figure}[t]
\includegraphics[width=0.6\textwidth,angle=270]{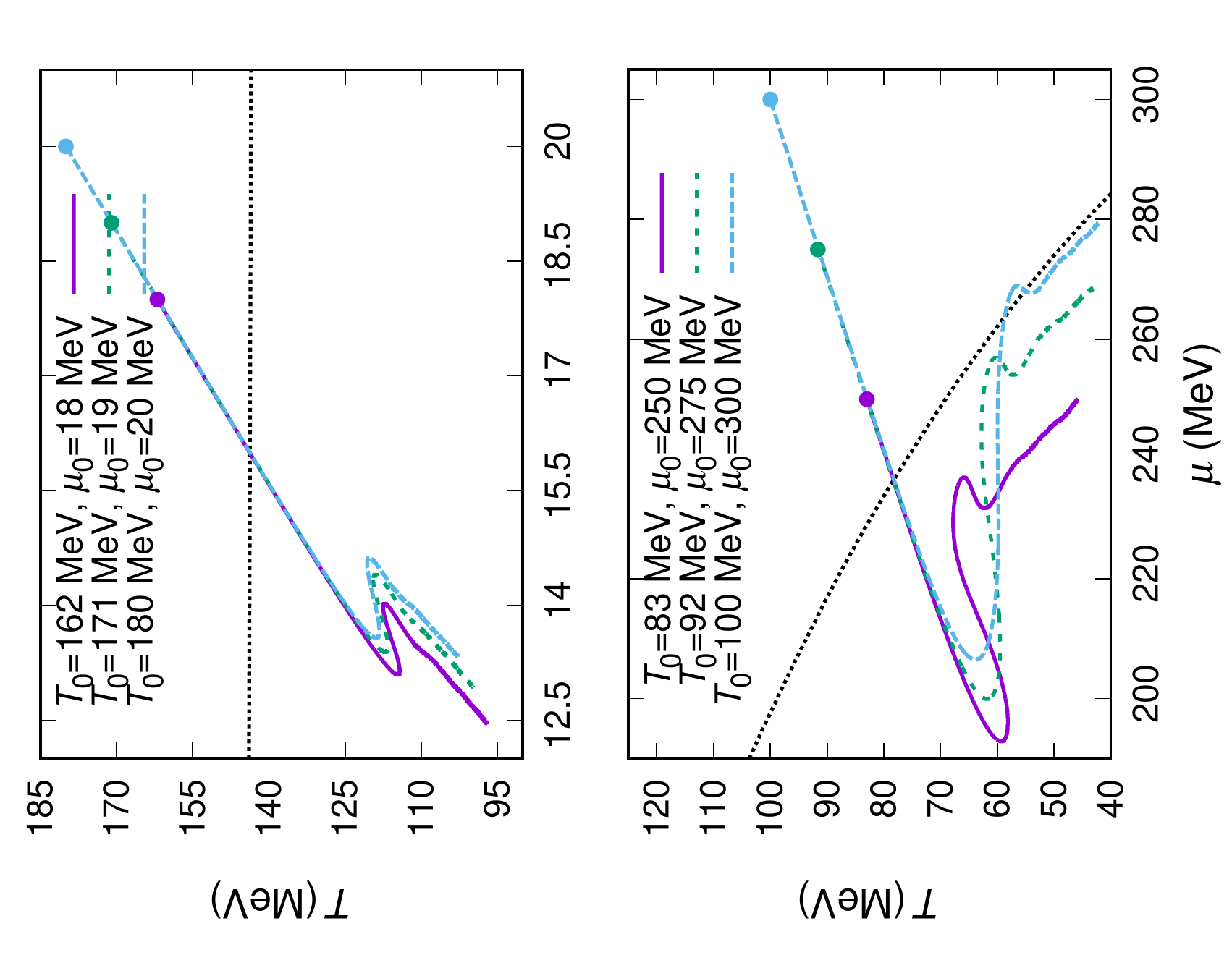}
\caption{\label{fig:traj_cr_fo} Top: Event-averaged trajectories for a crossover transition, starting on the same isentrope but with different distances to the phase boundary. Bottom: Event-averaged trajectories for a first-order phase transition. The dotted line delineates the crossover and first-order phase transition, respectively. } 
\end{figure} 

\begin{figure}[t]
\includegraphics[width=0.45\textwidth,angle=270]{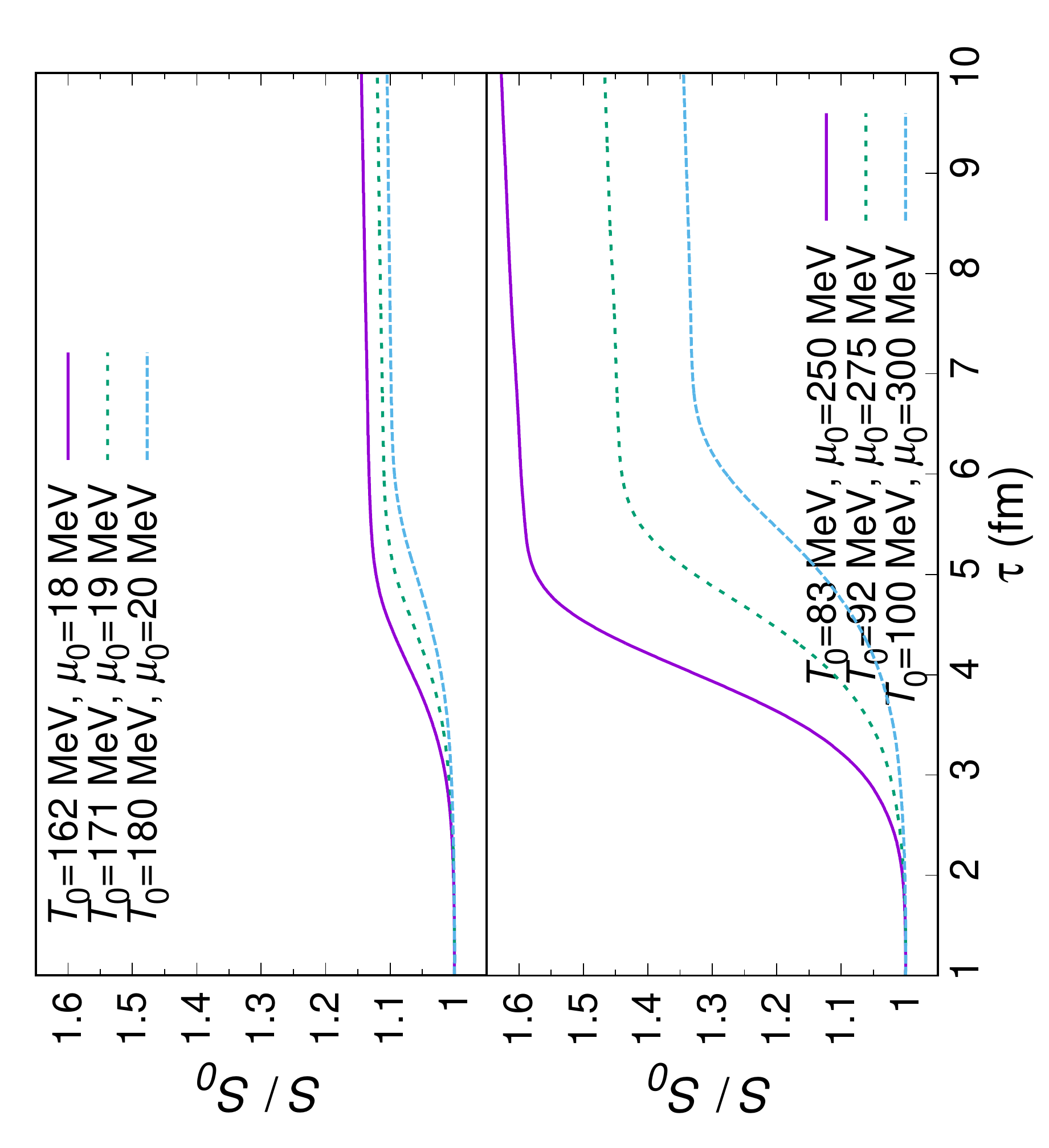}
\caption{\label{fig:s_cr_fo} Top: Event-averaged entropy increase for a crossover scenario, corresponding to the trajectories in Fig. \ref{fig:traj_cr_fo} (top). Bottom: Event-averaged entropy increase for a first-order phase transition, corresponding to the trajectories in Fig. \ref{fig:traj_cr_fo} (bottom).}
\end{figure} 

In this section, we are going to investigate the relation between the initial state and the amount of reheating at a crossover and first-order phase transition. We will see that this also has an effect on the evolution of the entropy. We choose three different initial conditions along an isentropic trajectory above the crossover region and the first-order phase transition line, each. The trajectories for these events are shown in two plots in Fig. \ref{fig:traj_cr_fo}. The first remarkable thing to observe is a reheating process after passing through the crossover, an effect that usually occurs after the decay of a supercooled state in a first-order phase transition. Although no supercooling happens for the dynamical crossover, a delay in the relaxation process during the rapid phase change will nevertheless lead to significant energy dissipation and consequently to an increase in $T$. We see that the trajectory below the crossover line depends on the initial conditions, trajectories starting closer to the crossover line tend to overshoot the phase boundary further. The reason for this is that for those events, the expansion rate $\sim 1/\tau$ at the phase boundary is larger, i.e.\ while the relaxation time is roughly the same, the more rapid expansion leads to a lower $T$ and $\mu$ before the relaxation process to the chirally broken phase starts. At a first-order phase transition, the overshooting effect is the same and has also been found before in inhomogeneous media \cite{Herold:2014zoa}. In contrast to the crossover, however, we can see a clearer difference in the evolution below the phase transition line. Starting closer to the phase boundary leads to a stronger reheating process which can here be ascribed to the formation and decay of a supercooled state.  The longer this state survives, the larger amount of energy will be dissipated into the fluid causing a larger rise in the temperature. In the context of heavy-ion experiments, it is important to understand this effect as it will create additional thermal background. Finally, we note that significantly higher values of $\mu$ are reached by choosing an initial state farther away from the first-order phase transition line. 

Fig. \ref{fig:s_cr_fo} shows the evolution of $S/S_0$. Here, we consider the relative increase in entropy by dividing through the initial entropy of the medium $S_0\equiv S(\tau=1$~fm$)$. We see that, in general, the entropy increases stronger at a first-order phase transition than at a crossover. Furthermore, the amount of increase depends on how close the initial condition is to the phase boundary, with a higher expansion rate resulting in a larger increase in $S$.

\section{Entropy increase with and without latent heat}
\label{sec:entropy}

\begin{figure}[t]
\includegraphics[width=0.35\textwidth,angle=270]{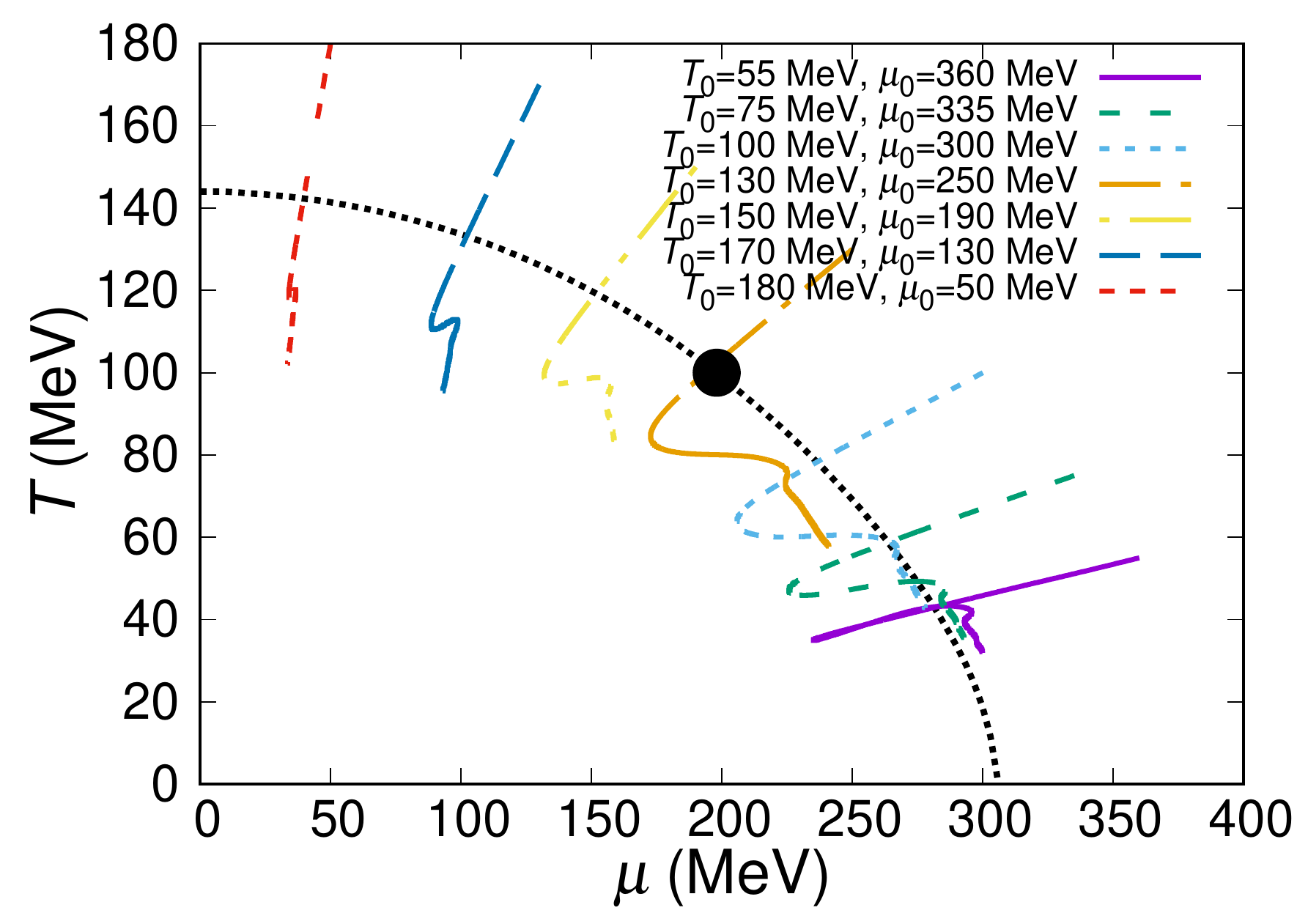}
\caption{\label{fig:traj} Event-averaged trajectories for several initial conditions, probing different regions of the phase diagram. The dashed line corresponds to the phase boundary and the black dot indicates the position of the CP.} 
\end{figure} 

\begin{figure}[t]
\includegraphics[width=0.35\textwidth,angle=270]{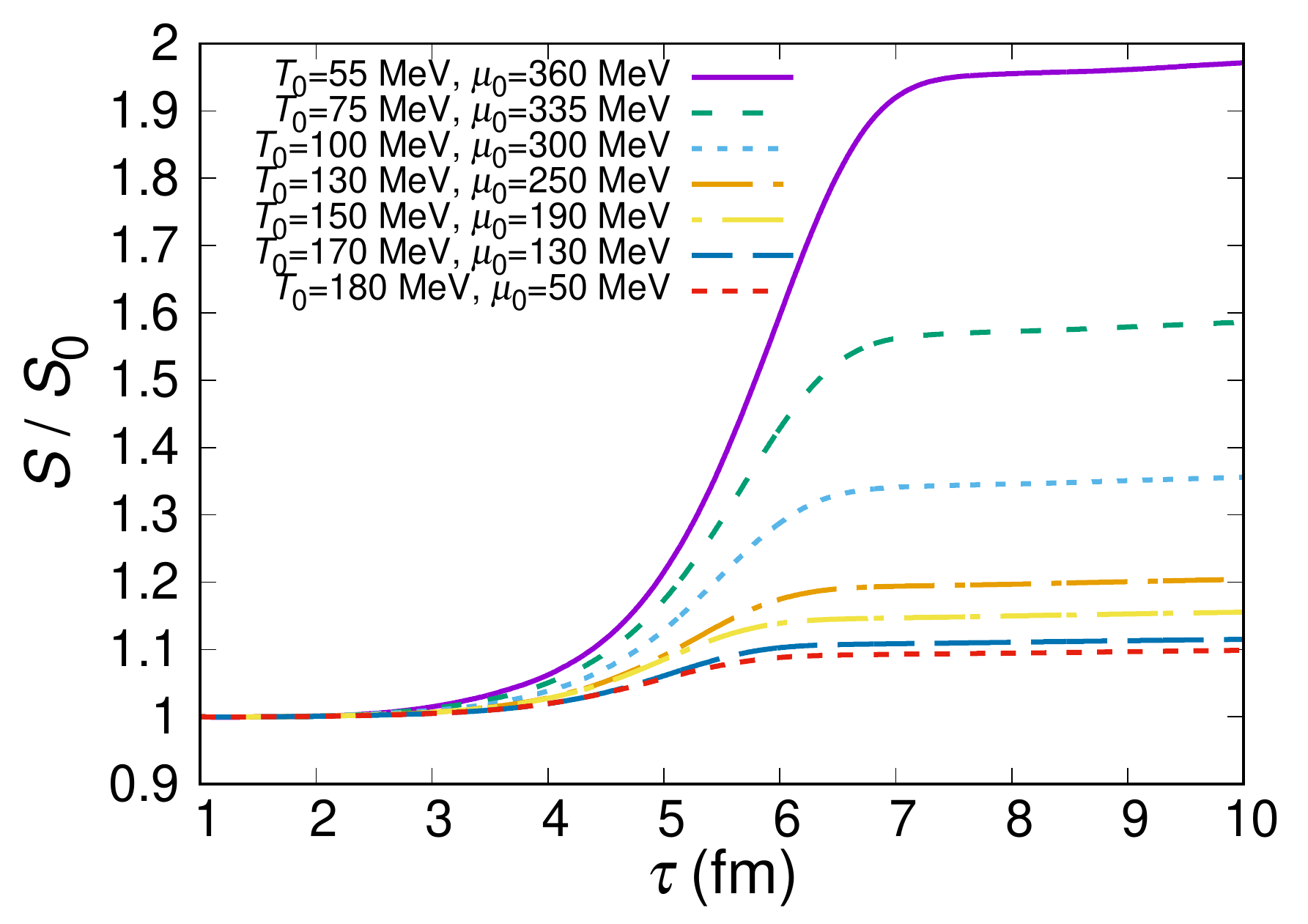}
\caption{\label{fig:s} Relative entropy increase for the trajectories shown in Fig. \ref{fig:traj}. We see that with increasing initial $\mu$, the entropy increase becomes more significant.} 
\end{figure} 

We are now prepared to estimate the entropy production for events with different initial densities, probing different regions of the phase diagram. To make the results comparable, we are choosing initial conditions all across the phase diagram in such a way that each trajectory is going to meet the phase boundary at a fixed proper time of $\tau=2$~fm. Although this does not necessarily reflect the experimental reality where the initial condition is solely determined by the beam energy and nucleon number, this will help us disentangle effects of different transition types from the impact of the expansion rate discussed in the previous section. The trajectories are shown in Fig. \ref{fig:traj}, together with the corresponding evolution of $S/S_0$ in Fig. \ref{fig:s}. A clear trend is found: While for a crossover and CP transition the entropy increase is of the order of $10-20\%$, slowly becoming larger for transitions closer to the CP, the presence of a latent heat amplifies this effect, resulting in an increase of up to $100\%$ for the trajectory with the strongest first-order phase transition. 

Experimentally, this effect can be expected to lead to an increase of the pion-to-baryon-number ratio, we therefore propose future experiments to search for steps in the $\pi/p$ multiplicity ratio as function of the beam energy.

\section{Conclusions}
\label{sec:conclusions}

We have studied the dynamical evolution of the zero mode of the chiral order parameter $\sigma$ during the longitudinal expansion of a hot and dense fluid, described by the coupling to Bjorken hydrodynamics. The results demonstrate that entropy is generated by damping processes during the transition and by stochastic fluctuations during the whole evolution. Interestingly, a reheating effect of the medium is observed not only for a first-order phase transition but also for events with a crossover. In all cases, the amount of reheating and entropy production depends on the expansion rate at the moment when the medium reaches the phase boundary. Assuming a proper time interval of $\sim 1$~fm between the thermalization of the medium and the start of the transition process, we were able to estimate the relative increase of $S/N$ ranging from $25\%$ for a crossover to up to $200\%$ for a first-order phase transition. 

As the present study is a very simple model for the production of entropy, more realistic approaches should be pursued in the future. Possible extensions of this work should consider an inhomogeneous medium, i.e.\ a full (3+1)-dimensional hydrodynamic expansion and include the effect of spatial fluctuations or higher-order modes of the chiral order parameter field.

\section*{Acknowledgments}
This work was supported by Suranaree University of Technology (SUT), TRF-RGJ, Deutscher Akademischer Austausch Dienst (DAAD), HIC for FAIR and in the framework of COST Action CA15213 THOR. C.H. and C.K. acknowledge support from SUT-CHE-NRU (FtR.15/2559) project. M.N. acknowledges the support of the program ``Etoiles montantes en Pays de la Loire 2017''.

\bibliographystyle{elsarticle-num}
\bibliography{mybib.bib}

\end{document}